\documentclass{aa}
\usepackage{times}
\usepackage{graphicx}
\begin{document}

%\thesaurus{02.04.01, 08.14.1, 08.16.6, 02.07.01}
\title{Accelerated expansion of the Crab Nebula and evaluation of 
its neutron-star parameters}

\author{M. Bejger \and P. Haensel}
 \institute{N. Copernicus Astronomical Center, Polish
       Academy of Sciences, Bartycka 18, PL-00-716 Warszawa, Poland\\
~~e-mail: {\tt  bejger@camk.edu.pl,\ haensel@camk.edu.pl
  }}
\date{
Received  xx xxxx, 2002/Accepted xx xxxx
}
\abstract{
A model of an accelerated expansion of the Crab Nebula powered by 
the spinning-down Crab pulsar is proposed, in which time dependence 
of the acceleration is connected with evolution of pulsar luminosity. 
Using recent observational data, we derive estimates of the 
Crab neutron-star moment of inertia. Correlations 
between the neutron star moment of inertia and its mass and radius 
allow for rough estimates of the Crab neutron-star radius and mass. 
In contrast to the previously used constant-acceleration approximation, 
even for the expanding nebula mass $\sim 7~M_{\odot}$ results obtained
within our model do not stay in conflict with the modern stiff equations 
of state of dense matter.
\keywords{neutron stars, moment of inertia, Crab pulsar}
}
\titlerunning{Accelerated expansion of the Crab Nebula}
\authorrunning{M. Bejger and P. Haensel}
\maketitle
\section{Introduction}
The AD 1054 supernova remnant, Crab Nebula, is probably the most often observed 
object in the sky. Optical observations of its filaments made in the past
century are sufficient to indicate that the 
motion of filaments is accelerated, ${\dot v} >0$. This accelerated expansion,
connected with the local interstellar medium sweeping, as well as the nebula 
radiation, are all powered by the Crab pulsar which was discovered in the center of 
the nebula in 1968. The energy reservoir is constituted by the pulsar rotational 
energy, which loses it at a rate ${\dot E}_{\rm rot}= I\Omega{\dot \Omega}$, where 
$I$ is the pulsar moment of 
inertia and $\Omega$ and $\dot\Omega$ are angular frequency and its time derivative,
both obtained from the pulsar timing. Assuming the balance between 
${\dot E}_{\rm rot}$ and the power of the nebula radiation and accelerated expansion 
in the interstellar medium, one gets a constraint on $I$, which in turn may be used to 
put a condition on the largely unknown equation of state (EOS) of dense matter.

Classical analysis along these lines was proposed and carried out by Manchester \&
Taylor (1977). Some thirteen years later, it was carried out by one of us using more
recent data on the Crab Nebula (Haensel 1990). In both cases, it has been assumed that
${\dot v}=const.$ during nebula expansion. Constraints derived by Manchester \& Taylor
(1977) were weak and did not eliminate any of EOSs. The later analysis in (Haensel
1990) pointed out crucial dependence on the mass of the expanding nebula $M_{\rm neb}$.
The highest of the estimates of $M_{\rm neb}$ available in 1980s ruled out the 
softest EOSs.

The most recent estimates of the mass contained in the optical filaments are
significantly higher than the previous ones ($4.6\pm 1.8~{\rm M}_\odot$, Fesen et al.
1997). As we have recently shown, putting  $M_{\rm neb}=4.6~{\rm M}_\odot$ in the
classical ${\dot v}=const.$ expansion model eliminates nearly all existing EOSs except
the stiffest ones (Bejger \& Haensel 2002). Actually, the situation can be even worse:
elementary model of type II supernovae predicts that a neutron star is a byproduct of
explosion of an evolved star with mass $\ga 8~{\rm M}_\odot$. Matter seen as filaments
constitutes only a part of eject mass, and with $M_{\rm neb}\sim 7~{\rm M}_\odot$ no
realistic EOS can provide Crab pulsar with sufficiently high $I$ to account for needed
${\dot E}_{\rm rot}$. This would eliminate all existing realistic EOSs of dense
matter.

Here we present a model of the Crab Nebula expansion which avoids
the artificial approximation ${\dot v}=const.$ and is consistent with stiff EOSs even
for $M_{\rm neb}\sim 7~{\rm M}_\odot$. We use ${\dot v}$ averaged in time, using a
standard model of the pulsar frequency evolution. This assumption, based on elementary pulsar
astrophysics, removes most of the drastic problems connected with high $M_{\rm neb}$.

In Sec. \ref{sect:obs.Ecomp} we summarize observational facts and apply them to the
description of the kinematics and energy budget of the Crab Nebula.
In Sec. \ref{sect:vdot.const} we briefly summarize results obtained using
the $\dot{v}=const.$ approximation. Our model for the
accelerated expansion is presented in Sec.\ \ref{sect:vdot.t-dep}. It is used to evaluate
$I$ of the Crab pulsar, which is then applied in Sect.\ \ref{sect:discussion}
to derive constraints on the dense matter EOS. Finally, we apply recently
derived formulae expressing $I$ in terms of the stellar mass and radius (Bejger \& Haensel
2002) to get constraints in the mass-radius plane for the neutron star and strange star
model of the Crab pulsar.
\newpage
%%%%%%%%%%%%%%%%%%%%%%%%%%%%%%%%%%%%%%%%%%%%%%%%%%%%%%%%%%%%%%%%%%%%%%%%%%%
\section{Observational facts and energy balance of the pulsar-nebula system}
\label{sect:obs.Ecomp}
%%%%%%%%%%%%%%%%%%%%%%%%%%%%%%%%%%%%%%%%%%%%%%%%%%%%%%%%%%%%%%%%%%%%%%%%%%%
Presently measured pulse period and the period
derivative of the Crab pulsar are $P_{\rm p}=0.0334033\ \rm s$ and 
$\dot{P}_{\rm p}=4.20996 \times  10^{-13}\ \rm{s~s^{-1}}$
(Taylor et al. 1993), which corresponds to the angular
frequency $\Omega_{\rm p}=188.101\ \rm s^{-1}$ and 
$\dot{\Omega}_{\rm p}=-2.37071 \times  10^{-9}\ \rm s^{-2}$.
The rotational energy of the neutron star is dissipated via the 
emission of particles, electromagnetic waves and through the interaction
of the pulsar with the surrounding gas.
 The value of ${\dot \Omega}$ can be related to $\Omega$ by
 %%%%%%%%%%%%%%%
\begin{equation}
\dot{\Omega}=-K \Omega^{n},
\label{omega_dot}
\end{equation}
%%%%%%%%%%%%%%%%%%
where $K$ and $n$ are constants to be determined from the pulsar timing.
In the c.g.s. units  $K=4.66 \times 10^{-15}$. The breaking index $n$ can be
expressed in terms of the measurable timing parameters $\Omega$, $\dot\Omega$,
and $\ddot\Omega$, namely $n={\Omega{\ddot\Omega}/{\dot\Omega}^2}~$. 
Its value for the Crab pulsar, calculated using the 1982-1987 timing data, is
$n=2.509 \pm 0.001$ (Lyne et al. 1988). We make a standard assumption 
that $n$ depends only on the pulsar
magnetic field, whose configuration was fixed after formation of the pulsar 
(e.g., in less than a few months). In what follows we will count the pulsar
age from that moment. Integration of the Eq.\ (\ref{omega_dot}) from $t=0$
to $t=T=938$ yr (the reason for choosing  this value of $T$ will become
clear later) will give us the initial angular frequency $\Omega_{\rm i}$ and
initial period $P_{\rm i}$:
%%%%%%%%%%%%%%%%%%%%%%%
\begin{equation}
\Omega_{\rm i}=\left[\Omega_{\rm p}^{1- n} - KT(n - 1)\right]
^{1/(1-n)}= 325.757\ \rm s^{-1}, 
\label{omega_init}
\end{equation}
%%%%%%%%%%%%%%%%%%%%%%%%%
%%%%%%%%%%%%%%%%%%%%%%%%%%%%%%
\begin{equation}
P_{\rm i}={2 \pi\over \Omega_{\rm i}}=0.0192880\ \rm s. 
\label{p_init}
\end{equation}
%%%%%%%%%%%%%%%%%%%%%%%%%%%%%%%%%%%
The loss of the rotational energy can be written as
%%%%%%%%%%%%%%%%%%%%%%%%%%%%%%%%%%%%%%%%%%%%%%
\begin{equation}
\dot{E}_{\rm rot}= \frac{\rm d}{\rm{d} t}\left(\frac{1}{2}I
\Omega^2\right)= -I \Omega
|\dot{\Omega}|, 
\label{eng_rot}
\end{equation}
%%%%%%%%%%%%%%%%%%%%%%%%%%%%%%%%%%%%%%%%%%%%%%%%%%
where small contribution resulting from the dependence of 
$I$ on $\Omega$ (increase of
$I$ is quadratic in $P_{\rm ms}/P_{\rm p}$, where  the mass-shedding 
period $P_{\rm ms}\sim 1$ ms) was neglected. The rotational energy of the pulsar is
transformed into radiation luminosity ${\dot E}_{\rm rad}$  and  the 
energy needed to support accelerated  nebula expansion  in the 
surrounding interstellar medium ${\dot E}_{\rm exp}$.

In order to make further calculations feasible, we will introduce approximation of
spherical symmetry. In principle, deviations from spherical symmetry can  be
accounted for by introducing corrections via ``anisotropy factors'' in the final
results. For the time being, we have no sufficient observational information to
implement such a procedure, and we will restrict ourselves to the
spherically-symmetric model. Following Petersen (1998) we write the total radiated
energy per unit time as
%%%%%%%%%%%%%%%%%%%%%%%%%%%%%
\begin{equation}
\dot{E}_{\rm rad}(D)\simeq 1.25\cdot \left(\frac{D}
{D_{\rm DF}}\right)^2 \times10^{38}\
\rm erg~s^{-1}, 
\label{eng_rad}
\end{equation}
%%%%%%%%%%%%%%%%%%%%
where $D$ is the distance to the nebula in kpc. The value $D_{\rm DF}=1.83$ kpc comes
from the paper of  Davidson \& Fesen (1985). When calculating 
 $\dot{E}_{\rm exp}$, we should take into account the fact that the
nebula expands in the interstellar medium. We will approximate the nebula by an
expanding spherical shell of radius $R_{\rm neb}$. 
The  shell expansion velocity is then $v={\dot R}_{\rm neb}$. Expanding shell will 
increase its mass by sweeping the interstellar medium after accelerating it 
to its own velocity $v$. Therefore, the expression for ${\dot E}_{\rm exp}$ reads
%%%%%%%%%%%%%%%%%%%%%%%%%%%%%%%%%%
\begin{equation}
\dot{E}_{\rm exp}= \frac{d}{\rm d t}\left( \frac{1}{2}M_{\rm neb}v^2\right)=M_{\rm
neb}v\dot{v}+ \frac{1}{2}\dot{M}_{\rm neb}v^2~,  
\label{eng_exp}
\end{equation}
%%%%%%%%%%%%%%%%%%%%%
where $M_{\rm neb}$ is the mass of the nebula. 
A  mass of the interstellar hydrogen 
added  to the nebula per unit time during the expansion 
of nebula in the interstellar medium is 
%%%%%%%%%%%%%%%%%%%%%%%%%%%%%%%%%%%%%%%%%%%%%%%%%
\begin{equation}
\dot{M}_{\rm neb}=4\pi R^2_{\rm neb}n_{\rm H} m_{\rm H} v~, 
\label{mneb_dot}
\end{equation}
%%%%%%%%%%%%%%%%%%%%%%%%%%%%%%%%%%%%%%%%%%%%%%%%%%%%
where $m_{\rm H}$ denotes the hydrogen atom mass and $n_{\rm H}$  is  the number
density of hydrogen atoms in space around the nebula. 
For our computation we
use the canonical value $n_{\rm H}= 0.2\ \rm cm^{-3}$ from Manchester~\&~Taylor
(1977). 

Our spherical-shell model is a simplest possible representation of Crab 
Nebula, which is famous for its rather complicated crab-like shape. 
The value of $R_{\rm neb}$ will be evaluated as a mean for an ellipsoid 
which is a more precise model of the shape of the Crab Nebula.  
Assuming  $D=D_{\rm DF}$, one gets then $R_{\rm neb}=1.25\ \rm pc$ 
(see e.g. Douvion et al. 2001).

The present mass of the Crab Nebula, $M_{\rm neb}$, will play 
a central role in our model. Its observational estimation
is very difficult - in the last two decades  the value varied in time from
$2-3\ M_{\odot}$ (Davidson \& Fesen 1985), through $1-2\ M_{\odot}$ (MacAlpine
\& Uomoto 1991) to $4.6 \pm 1.8\ M_{\odot}$ (Fesen et al. 1997).

%%%%%%%%%%%%%%%%%%%%%%% Fig. 1 %%%%%%%%%%%%%%%%%%%%%%%%%%%%%%%%%%%%%%%%
\begin{figure}
\centering
\resizebox{3.25in}{!}{\includegraphics{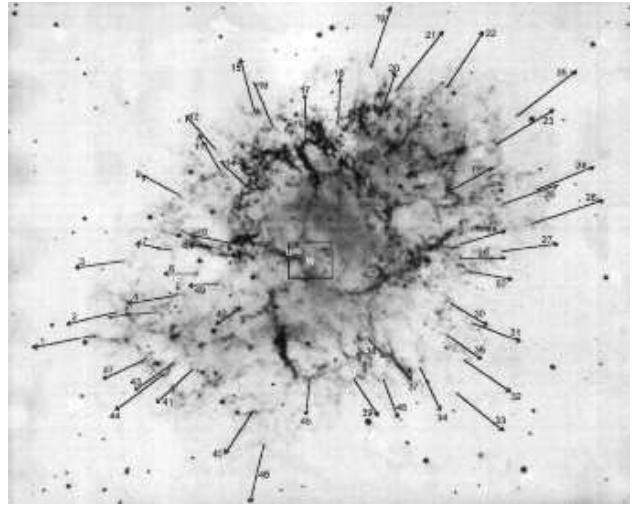}}
\caption{Expansion of the Crab Nebula. Arrows represent motions of 50 
optical filaments in next 250 yr
at current expansion rates. From Nugent (1998), with kind permission of
the author.}
\label{fig:expCrab}
\end{figure}
%%%%%%%%%%%%%%%%%%%%%%%%%%%%%%%%%%%%%%%%%%%%%%%%%%%%%%%%%%%%%%%%%%%%%%%

The  expanding nebula shell is filled with 
optically shining filaments, whose motion can be measured by comparing 
the filaments positions on the high-resolution photographs taken more 
than 15-20 years apart (Duncan 1939, Trimble 1968, Wyckoff \& Murray 1977, 
Nugent 1998). In the present paper we will use the most recent results 
obtained by Nugent (1998). By comparing positions of 50 identifiable 
bright filaments on high-resolution plates taken in 1939, 1960, 1976, 
and 1992, Nugent calculated the mean velocity of their expansion. His 
results are visualized in Fig. \ref{fig:expCrab}, 
which was for us a source of inspiration for studying the Crab Nebula 
dynamics. By projecting the 
straight-line  constant velocity motion of filaments backward in time, 
Nugent obtained convergence of filaments trajectories at AD $1130\pm 16$ yr. 
His result was in accordance with previous estimate of Trimble (1968). 
Had the nebula expanded at a constant $v$, this would be the moment of 
Crab supernova explosion. However, the date recorded by the Chinese 
astronomers is AD 1054, which is $\Delta=76$ yr earlier. Therefore, 
the expansion had a non-zero acceleration $\dot v$.  
During expansion, $v$ increased from initial $v_{\rm i}$ 
to the present $v_{\rm p}$, known also from the spectra measurements 
(e.g. Sollerman et al. 2000), $v_{\rm p} \sim 1.5 \times 10^8 {\rm cm/s}$. 
%%%%%%%%%%%%%%%%%%%%%%%%%%%%%%%%%%%%%%%%

%%%%%%%%%%%%%%%%%%%%%%%%%%%%%%%%%%%%%%%%%%%%%%%%%%%%%%%%%%%%%%%%%%%%%%%%
\section{Crab Nebula dynamics and bounds for the moment of inertia of its
         neutron star}
\label{sect:I.model.bounds}
%%%%%%%%%%%%%%%%%%%%%%%%%%%%%%%%%%%%%%%%%%%%%%%%%%%%%%%%%%%%%%%%%%%%%%%%
\subsection{Constant acceleration}
\label{sect:vdot.const}
%%%%%%%%%%%%%%%%%%%%%%%%%%%%%%%%%%%%%%%%%%
In view of the lack of information on the time dependence of $\dot v$ during 
the nebula lifetime, the most natural approximation is to consider it 
as constant in time. This is the approximation used in previous studies 
(Manchester \& Taylor 1977, Haensel 1990, Bejger \& Haensel 2002). 
This constant value of acceleration will be denoted by $\dot{v}_{\rm c}$, 
and can be calculated from the existing data using the formula 
%%%%%%%%%%%%%%%%%%%%%%%%%%%%%%%%%%%
\begin{equation}
\dot{v}_{\rm c}={2\Delta v_{\rm p}\over T^2}~, 
\label{acc_const}
\end{equation}
%%%%%%%%%%%%%%%%%%%%%%%%%%%%%%%%%%%%%%
where $T=938~{\rm yr}$ is the lifetime of the nebula from birth in 
1054 AD to Nugent's photographic evaluation in 1992. 
Putting numerical values, we get 
$\dot{v}_{\rm c}=0.82\times 10^{-3}\ \rm{cm~s^{-2}}$. 

The knowledge of the present $v$ and $\dot{v}$
allows one  to get expression for the Crab pulsar 
moment of inertia. This expression results from the condition that the loss of 
the kinetic rotational energy of the pulsar should be 
sufficient to support 
$\dot{E}_{\rm rad}+\dot{E}_{\rm exp}$, 
%%%%%%%%%%%%%%%%%%%
\begin{eqnarray}
I_{\rm Crab}&\ge& \left[
\dot{E}_{\rm rad}(D)+M_{\rm neb}v\dot{v}\right]/
(\Omega|\dot{\Omega}|)~
\cr\cr
&~&+2\pi R_{\rm neb}^2 n_{\rm H} m_{\rm H} v^3/
(\Omega |\dot{\Omega}|)~.
\label{eq_fin}
\end{eqnarray}
%%%%%%%%%%%%%%%%%%%%%
The above equation is generally valid and does not involve 
assumption on a time dependence of $v$ and $\dot{v}$. 

From Eq. (\ref{eq_fin}), using $M_{\rm neb}=4.6\ M_{\odot}$, $D=1.83$ kpc,
$v=v_{\rm p}=1.5\times 10^{8}~{\rm cm~s^{-1}}$, $\dot{v}=\dot{v}_{\rm c}=0.82\times 
10^{-3}~{\rm cm~s^{-2}}$ and
$\Omega_{\rm p} |\dot{\Omega}_{\rm p}|=4.459\times 10^{-7}\ \rm s^{-3}$
we get an estimate of a lower bound on $I_{\rm Crab}$, 
labeled with ``c'' which reminds 
the constant acceleration assumption,
%%%%%%%%%%%%%%%%%%%%%%%%
\begin{eqnarray}
I_{\rm Crab,45}&\ge&I^{\rm (c)}_{45}= 0.28 \left({D\over D_{\rm DF}}\right)^2
+ 2.53 {M_{\rm neb}\over 4.6~M_\odot}\cr\cr
&~& +0.23 \left({R_{\rm neb}\over 1.25~{\rm pc}}\right)^2 {n_{\rm H}\over
0.2~{\rm cm^{-3}} }~. 
\label{Ic.old}
\end{eqnarray}
%%%%%%%%%%%%%%%%%%%%%%%%%%%%%%%%%%%%%%%%%%%%%%%%%%%%%%
where $I_{45}\equiv I/10^{45}~{\rm g~cm^2}$. 
With our choice of parameters, this equation  yields  $I^{\rm (c)}_{45}=3.04$. 
As shown in our previous paper (Bejger \& Haensel 2002), such a value of 
$I_{\rm Crab}$ requires a very stiff EOS of dense matter. Therefore, 
the constraint on the maximum moment of inertia for a dense matter 
EOS, $I_{\rm max}$, which should satisfy  the inequality 
$I_{\rm max}>I_{\rm Crab}$, is very strong. 
Actually, the problem can become quite dramatic if the 
expanding shell contains a typical mass ejected in a type II supernova, 
because for $M_{\rm neb}\ga 7~M_\odot$ all existing realistic EOSs are 
ruled out by the $I_{\rm max}>I_{\rm Crab}$ condition. However, as 
we show in the next section this may 
just result from the unrealistic  character of the
assumption $\dot{v}=const.$ 
%%%%%%%%%%%%%%%%%%%%%%%%%%%%%%%%%%%%%%%%%%%%%%%%%%%%%%%%%%%%
\subsection{Time-dependent acceleration}
\label{sect:vdot.t-dep}
%%%%%%%%%%%%%%%%%%%%%%%%%%%%%%%%%%%%%%%%%%%%%%%%%%%%%%%%%%
As we can see, the acceleration term is largely dominating in 
the r.h.s. of Eq. (\ref{Ic.old}).
We will assume that this dominance was valid also in the past,
after some initial short-term period ($<$3 yr) in which the Crab
Nebula was powered by the sources connected with supernova 
itself (i.e., $^{56}{\rm Ni}$ radioactive decay heating).  
Therefore, the loss of the pulsar rotational kinetic energy goes 
mainly into accelerating the nebula,
%%%%%%%%%%%%%%%%%%
\begin{equation}
I\Omega|\dot{\Omega}|\simeq M_{\rm neb}v\dot{v}~.
\label{vvdot.eq1}
\end{equation}
%%%%%%%%%%%%%%%%%%%%%%
As we argued before, at $P>P_{\rm i}=19$ ms the dependence  of $I$ on 
$P$ (and therefore on time) is negligible. On the other hand, total 
increase of the nebula mass since 1054 due to the sweeping of the 
interstellar medium can be estimated as 
%%%%%%%%%%%%%%%%%%%%%%%
\begin{equation}
\Delta M_{\rm neb}={4\over 3} \pi R_{\rm neb}^3 n_{\rm H} m_{\rm H}= 
0.04~M_\odot~.
\label{DeltaMneb}
\end{equation}
%%%%%%%%%%%%%%%%%%%%%%%%%%
This increase can be neglected compared to the present nebula mass,
so that the assumption $M_{\rm neb}\simeq const.$ is valid.
Using Eq. (\ref{vvdot.eq1}) 
we can therefore approximately express $v\dot{v}$ at any moment in the past as 
%%%%%%%%%%%%%%%%%%%%%%%%
\begin{equation}
v\dot{v}\simeq {I\Omega|\dot{\Omega}|\over M_{\rm neb}}~.
\label{vvdot.eq2}
\end{equation}
%%%%%%%%%%%%%%%%%%
During 938 yr of expansion, $v$ increased by some $2\Delta/T\sim 16\%$, 
(assuming $\dot{v}=const.$, see previous subsection) 
which is a relatively small change compared to the change in 
$\Omega|\dot{\Omega}|$. 
Namely, equation 
%%%%%%%%%%%%%%%%%
\begin{equation}
\Omega|\dot{\Omega}|=K\Omega^{n+1}=K\Omega^{3.509}~
\label{Omega.Omegadot}
\end{equation}
%%%%%%%%%%%%%%%%%%%%
implies that during nebula lifetime the present 
$\Omega|\dot{\Omega}|$ decreased by a factor 
%%%%%%%%%%%%%%%%%%%%%%%%%
\begin{equation}
\left({\Omega_{\rm i}\over \Omega_{\rm p}}\right)^{3.509} 
=\left({P_{\rm p}\over P_{\rm i}}\right)^{3.509}=6.84.
\label{decrease.Omega.Omegadot}
\end{equation}
%%%%%%%%%%%%%%%%%%%%%%%%%%%
This indicates a rather strong dependence of the product 
$v\dot{v}$ on time. However, in view of a rather small
 increase in $v$, the time dependence of $v\dot{v}$
results mainly from a strong {\it decrease} of $\dot{v}$, 
%%%%%%%%%%%%%%%%%%%
\begin{equation}
\dot{v}(t)\simeq 
{I\Omega|\dot{\Omega}|\over M_{\rm neb} v_{\rm av}},
\label{vdot.form.t}
\end{equation}
%%%%%%%%%%%%%%%%%%%%%%%%%%
where we approximated $v$ by its time-averaged value $v_{\rm av}=R_{\rm neb}/T$. 
The average value of the acceleration can be calculated from 
%%%%%%%%%%%%%%%%%%%%
\begin{equation}
\dot{v}_{\rm av}=\frac{1}{T}{\int}_{0}^{T} \dot{v}\rm{d}t.
\label{acc_aver}
\end{equation}
%%%%%%%%%%%%%%%%%%%%%%%%%%%
Using previously derived expressions, we get 
%%%%%%%%%%%%%%%%%%%%%%%%%%%%%%%
%%%%%%%%%%%%%%%%%%%%%%%%%%%%%%%%%%%%
\begin{equation}
\dot{v}_{\rm av}=\frac{I}{MT_{\rm neb}v_{\rm av}}
{\int}_{\Omega_{\rm i}}
^{\Omega_{\rm p}}\Omega|\dot{\Omega}|{\rm d}t
=\frac{I}{2TM_{\rm neb}v_{\rm av}}(\Omega_{\rm i}^2 - \Omega_{\rm p}^2).
\label{vdot.av}
\end{equation}
%%%%%%%%%%%%%%%%%%%%%%%%%%
On the other hand, in the constant-acceleration model 
one would get 
%%%%%%%%%%%%%%%%%%%%%
\begin{equation}
I\Omega_{\rm p}|\dot{\Omega}_{\rm p}|\simeq 
M_{\rm neb}v_{\rm av}\dot{v}_{\rm c}.
\label{vdot.const.rel}
\end{equation}
%%%%%%%%%%%%%%%%%%%%%%%%%%%%
Therefore, 
%%%%%%%%%%%%%%%%%%%%%%%%%%
\begin{equation}
{\dot{v}_{\rm av}\over \dot{v}_{\rm c}}\simeq 
{\Omega_{\rm i}^2 - \Omega_{\rm p}^2\over 
2 T \Omega_{\rm p}|\dot{\Omega}_{\rm p}|}=2.67.
\label{vdot.decrease}
\end{equation}
%%%%%%%%%%%%%%%%%%%%%%%%
The above result indicates that under conditions prevailing 
during the Crab Nebula expansion the assumption $\dot{v}=const.$ 
is not valid. 
In what follows, we will use the approximation
$I\Omega\dot{\Omega}\simeq M_{\rm neb}v\dot{v}$.

Within our model we can determine the present value of acceleration 
of the nebula expansion, using the method described below. We start with 
an elementary formula
%%%%%%%%%%%%%%
\begin{equation}
v_{\rm p}=v_{\rm i} +\int_0^T \dot{v}{\rm d}t~.
\label{vp.vi}
\end{equation}
%%%%%%%%%%%%%%%%%
Another elementary relation determines the average speed of expansion, 
%%%%%%%%%%%%%
\begin{equation}
v_{\rm av}={1\over T}{\int_0^T v {\rm d}t} = {R_{\rm neb}\over T} 
= 1.3 \times 10^8~{\rm cm~s^{-1}}.
\label{vav.int.T}
\end{equation}
%%%%%%%%%%%%%%%%
Changing the variable $t$ into $\Omega$, and carrying out the integrations 
over $\Omega$, we finally get a system of two equations for 
$v_{\rm i}$ and for $I$. The explicit form of this 
system of equations is: 
%%%%%%%%%%%%%%%%%%%%%%%%
\begin{eqnarray}
&v_{\rm p} & =  v_{\rm i} + \frac{I\left(\Omega_{\rm i}^2 - 
\Omega_{\rm p}^2\right)}{2M_{\rm neb}v_{\rm av}}, \nonumber \\
&R_{\rm neb} & = v_{\rm i}T + {I\Omega_{\rm i}^{3-n}\over
2(3-n)(1-n)KM_{\rm neb}v_{\rm av}}\times \nonumber\\
&\times&\left(2 + \left({\Omega_{\rm p}}/{\Omega_{\rm i}}\right)^{3-n}
\left[1-n - (3-n)\left({\Omega_{\rm i}}/{\Omega_{\rm
p}}\right)^2\right]\right), 
\label{Eqs.vi.I0}
\end{eqnarray}
%%%%%%%%%%%%%%%%%%
where the Crab pulsar timing constants $n$ and $K$ are given in 
Sect. \ref{sect:obs.Ecomp}.
As a result we get $v_{\rm i} = 0.93 \times 10^8~{\rm cm~s^{-1}}$, 
and for an assumed $M_{\rm neb}$ we are thus able to calculate
the value of $I_{\rm Crab}$.   
For the central and the upper value 
of $M_{\rm neb}$ obtained by Fesen et al. (1997) we get the following
numbers: 
%%%%%%%%%%%%%%%%%%
\begin{eqnarray}
M_{\rm neb}&=&4.6~M_\odot~~~\Longrightarrow~~~~I_{\rm Crab,45}>1.93~,\cr\cr
M_{\rm neb}&=&6.4~M_\odot~~~\Longrightarrow~~~~I_{\rm Crab,45}>2.68~.
\label{LB.ICrab}
\end{eqnarray}
%%%%%%%%%%%%%%%%%%
The value of $v_{\rm i}$ deserves a comment. It indicates that the
initial energy of expanding filaments-shell, 
%%%%%%%%%%%%%M
\begin{equation}
E^{\rm shell}_{\rm kin,i} = 4.6\times 10^{49}~\frac{M_{\rm
neb}}{4.6~M_{\odot}}\left(\frac{v_{\rm i}}
{10^8~{\rm cm~s^{-1}}}\right)^2~{\rm erg}
\label{shell.eq}
\end{equation}
%%%%%%%%%%%%%%%%
is much smaller that the canonical value expected for the core-collapse
supernovae, $10^{51}$ erg. Missing kinetic energy may reside in a 
fast-moving outer shell of the supernova remnant (Chevalier 1977).
Some evidence for the existence of such a fast-moving outer shell 
was found using the far-ultraviolet and optical {\it HST} observations
(Sollerman et al. 2000) 

%%%%%%%%%%%%%%%%%%%%%%%%%%%%%%%%%%%%%%%%%%%%%%%%%%%%%%%%%%%%%%%%%%%%%%%
%%%%%%%%%%%%%%%%%%%%%%% Fig. 1 %%%%%%%%%%%%%%%%%%%%%%%%%%%%%%%%%%%%%%%%
\begin{figure}
\resizebox{\hsize}{!}{\includegraphics{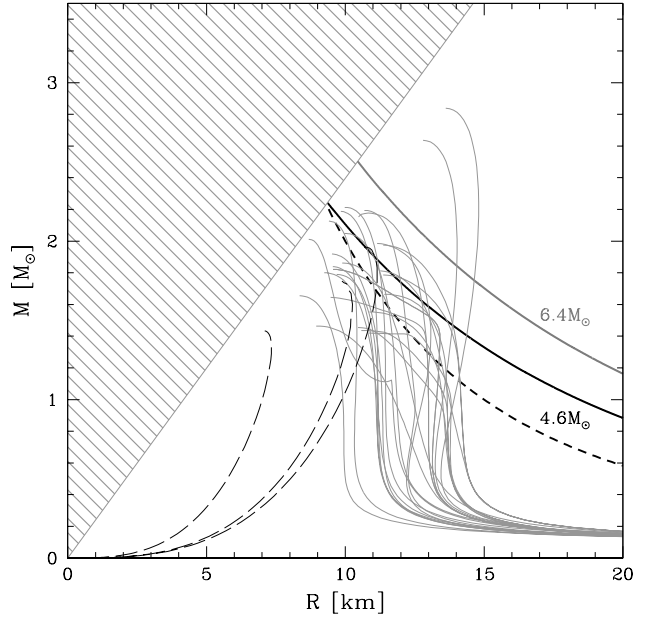}}
\caption{
The estimates for the moment of inertia of the Crab pulsar plotted
on the radius-mass diagram, when $M_{\rm neb}=4.6~M_{\odot}$, and 
$M_{\rm neb}=6.4~M_{\odot}$. 
Thin lines represents various equations of state, dotted - neutron stars,
long-dashed - strange stars. The set of thirty EOSs is the same as 
in Bejger~\&~Haensel (2002). In order to get constraints in $M-R$
plane, we use empirical $I(M,~R)$ relations
calculated by Bejger~\&~Haensel (2002) for neutron stars (thick solid
lines) and strange stars (thick dashed line). The shaded area is
excluded by General Relativity and the $v_{\rm sound}\le c$ condition.
}
\label{LB_Icrab}
\end{figure}
%%%%%%%%%%%%%%%%%%%%%%%%%%%%%%%%%%%%%%%%%%%%%%%%%%%%%%%%%%%%%%%%%%%%%%%

%%%%%%%%%%%%%%%%%%%%%%%%%%%%%%%%%%%%%%%%%%%%%%%%%%%%%%%%%%%%%%%%%%%%%%
\section{Constraints on the EOS, and $M$, and $R$ of the Crab pulsar}
\label{sect:discussion}
%%%%%%%%%%%%%%%%%%%%%%%%%%%%%%%%%%%%%%%%%%%%%%%%%%%%%%%%%%%%%%%%%%%%%%%%
Within a simple astrophysical model of the time-depending acceleration 
of the Crab Nebula expansion, 
we deduce constraints on dense matter EOS. These constraints depend on the 
mass of the Crab Nebula. 

For a central value obtained by Fesen et al. (1997) the constant-acceleration 
model one has $I_{\rm Crab,45}>3.04$ which could be allowed only by the 
stiffest EOSs with $M_{\rm max}>2~M_\odot$  (Bejger \& Haensel 2002). 
With time-dependent acceleration we get $I_{\rm Crab}$ 
which  is some 40\% lower, and this would exclude only soft EOSs  and 
those EOSs which are strongly softened at supra-nuclear density (due to the 
presence of hyperons or a phase transition). 
Within our model of $\dot{v}(t)$  the Crab 
pulsar can also power the nebula with uppermost value obtained by 
Fesen et al. (1997). We get then $I_{\rm Crab,45}(6.4~M_\odot)=2.68$, which 
leaves us with only very stiff EOS with $M_{\rm max}>2~M_\odot$. Even $M\sim 
7~M_\odot$ could be accommodated by existing stiff EOSs of matter composed 
of nucleons and leptons. Within constant-acceleration models, such values 
of $M_{\rm neb}$ rule out all existing realistic EOS of dense matter. 

Using  empirical in nature but actually very precise relation
between the moment of inertia, mass of the star and the corresponding radius
for neutron stars and strange quark stars 
(Bejger~\&~Haensel 2002) we plotted curves $I(M,~R)=I_{\rm Crab}$ in 
the mass-radius  diagram (Fig.\ \ref{LB_Icrab}). 
From this plot we deduce constraints on the mass and radius of Crab 
neutron star. If $M_{\rm neb}=4.6~M_\odot$ then neutron star has 
$M>1.5~M_\odot$ and $R=11-15~$km.  If the Crab pulsar is a strange star 
(a rather unlikely situation because of glitches, see Alpar 1987), then 
it has to have mass $M>1.7~M_\odot$ and $R=10-11~$km. If $M_{\rm neb}=6.4
~M_\odot$, then the EOS should be stiff, and we get $M>1.7~M_\odot$ 
and $R=12-15~$km. In the case of ``canonical theoretical'' $M_{\rm neb}\ga 
7~M_\odot$, the Crab neutron star is even more massive and lowest accepted $R$ 
is even larger. For such high masses $M_{\rm neb}\ga 6~M_\odot$ strange 
quark stars are ruled out.

%%%%%%%%%%%%%%%%%%%%%%%%%%%%%%%%%%%%%%%%%%%%%%%%%%%%%%%%%%%%%%%%%
\acknowledgements{ We would like to thank R. Nugent for the permission
to use Fig.1 in our paper. This work was partially supported 
by the KBN grants No. 5 P03D 020 20 and 2 P03D 004 22.}
%%%%%%%%%%%%%%%%%%%%%%%%%%%%%%%%%%%%%%%%%%%%%%%%%%

\end{document}